# Bilayer TeO₂: The First Oxide Semiconductor with Symmetric Sub-5-nm NMOS and PMOS

Linqiang Xu,[1,2,‡] Liya Zhao,[3,‡] Chit Siong Lau,[4] Pan Zhang,[3] Lianqiang Xu,[5] Qiuhui Li,[1] Shibo Fang,[1] Yee Sin Ang,[2,*] Xiaotian Sun,[3,*] and Jing Lu[1,6,7,8,9,*]

[1]State Key Laboratory of Mesoscopic Physics and Department of Physics, Peking University, Beijing 100871, P. R. China

[2]Science, Mathematics and Technology, Singapore University of Technology and Design (SUTD), 8 Somapah Road, Singapore 487372, Singapore

[3]College of Chemistry and Chemical Engineering, and Henan Key Laboratory of Function-Oriented Porous Materials, Luoyang Normal University, Luoyang 471934, P. R. China

[4]Institute of Materials Research and Engineering (IMRE), Agency for Science, Technology and Research (A*STAR), Singapore 138634, Republic of Singapore

[5]School of Physics and Electronic Information Engineering, Engineering Research Center of Nanostructure and Functional Materials, Ningxia Normal University, Guyuan 756000, China

[6]Collaborative Innovation Center of Quantum Matter, Beijing 100871, P. R. China

[7]Beijing Key Laboratory for Magnetoelectric Materials and Devices, Beijing 100871, P. R. China

[8]Peking University Yangtze Delta Institute of Optoelectronics, Nantong 226000, P. R. China

[9]Key Laboratory for the Physics and Chemistry of Nanodevices, Peking University, Beijing 100871, P. R. China

*Corresponding Authors: yeesin_ang@sutd.edu.sg, sxt@lynu.edu.cn, jinglu@pku.edu.cn

‡These authors contribute equally to this work.

## Abstract

Wide bandgap oxide semiconductors are very promising channel candidates for next-generation electronics due to their large-area manufacturing, high-quality dielectrics, low contact resistance, and low leakage current. However, the absence of ultra-short gate length ($L_g$) $p$-type transistors has restricted their application in future complementary metal-oxide-semiconductor (CMOS) integration. Inspired by the successfully grown high-hole mobility bilayer (BL) beta tellurium dioxide ($β$-TeO₂), we investigate the performance of sub-5-nm-$L_g$ BL $β$-TeO₂ field-effect transistors (FETs) by utilizing first-principles quantum transport simulation. The distinctive anisotropy of BL $β$-TeO₂ yields different transport properties. In the $y$-direction, both the sub-5-nm-$L_g$ $n$-type and $p$-type BL $β$-TeO₂ FETs can fulfill the International Technology Roadmap for Semiconductors (ITRS) criteria for high-performance (HP) devices, which are superior to the reported oxide FETs (only $n$-type). Remarkably, we for the first time demonstrate the existence of the NMOS and PMOS symmetry in sub-5-nm-$L_g$ oxide semiconductor FETs. As to the $x$-direction, the $n$-type BL $β$-TeO₂ FETs satisfy both the ITRS HP and low-power (LP) requirements with $L_g$ down to 3 nm. Consequently, our work shed light on the tremendous prospects of BL $β$-TeO₂ for CMOS application.

# 1. Introduction

Complementary metal-oxide-semiconductor (CMOS), which consists of NMOS and PMOS, plays a key role in state-of-art integrated circuits.[1-3] Over the past few decades, the development of CMOS technology has been driven by the scaling of field-effect transistors (FETs). However, contemporary silicon (Si)-based FETs suffer from severe short-channel effects (SCEs) when the gate length ($L_g$) is less than 10 nm .[4, 5] Emerging two-dimensional (2D) semiconductors offer promise as alternative channel materials because their atomic thickness is beneficial for mitigating SCEs and enhancing gate controllability.[5-8] Nevertheless, three primary challenges limited the further application of 2D semiconductors. Firstly, the large-area growth of 2D semiconductors compatible with CMOS technology remains a formidable obstacle.[6, 9] Secondly, it is very difficult to form high-quality dielectrics on the smooth surface of 2D semiconductors.[6, 10-12] Lastly, achieving proper doping of 2D semiconductors remains a significant challenge. An extra metal electrode is usually necessary to inject carriers into the 2D channel, which inevitably results in high contact resistance due to the Schottky barrier at the metal-semiconductor interface.[13-16]

Wide bandgap oxide semiconductors such as indium−gallium−zinc-oxide (IGZO), the channel material of thin-film transistors (TFTs), have been widely applied in the flat-panel display area.[17-19] The large-scale manufacturing of oxide semiconductors is a commercially mature technology, rendering them more appealing than 2D semiconductors for CMOS integration. Oxide semiconductors offer another advantage over 2D semiconductors, which is the ease of forming high-quality dielectrics on their surface via the atomic layer deposition (ALD) method.[10, 20, 21] This is because the dangling-bond surface of a 3D material is conductive to the deposition of dielectrics. Furthermore, high-carrier doping and, thus, low resistance can be easily realized in oxide semiconductors. For example, indium tin oxide (ITO) possesses an exceptionally low resistivity of about $10^{-4}$ Ω·cm, which is attributed to oxygen vacancy and substitutional tin dopants.[22] In contrast to Si, the oxide semiconductors are more suitable for low-power (LP) electronics because their wider bandgap (usually > 3 eV for oxide semiconductors *vs* 1.12 eV for Si) leads to a smaller leakage current.[11, 17] Consequently, these advantages position oxide semiconductors as competitive channel candidates for next-generation CMOS applications.

Extensive investigations have been conducted on the oxide semiconductor MOSFETs, primarily focusing on long-channel devices.[18, 23] Recently, with the thickness ($t$) thinned down to the nanometer scale, oxide semiconductors have emerged as promising candidates for ultra-short scale (sub-10 nm $L_g$) electronic devices such as ultrathin $In_2O_3$, monolayer (ML) $Ga_2O_3$, and bilayer (BL) $Ga_2O_3$ systems.[20, 24-27] However, all the reported ultra-short oxide semiconductors MOSFETs so far are only suitable for *n*-type devices in terms of the International Technology Roadmap for Semiconductors (ITRS) requirements.[28] The lack of *p*-type ultra-short oxide semiconductor transistors has hindered its further application in scaled CMOS. More importantly, whether the symmetric *n*- and *p*-type FETs can be achieved in such small oxide devices remains unknown.

Experimentally, the grown 2D *β*-TeO$_2$ with *t* down to BL is identified as a promising *p*-type oxide semiconductor, which exhibits high hole mobility of 232 cm$^2$·V$^{-1}$·s$^{-1}$ at room temperature.[29] We select the thinnest *β*-TeO$_2$ (BL *β*-TeO$_2$) as the channel material, and the transport characteristics of corresponding sub-5-nm-$L_g$ MOSFETs are simulated by the *ab initio* quantum transport simulation. The highly anisotropic electronic structures of BL *β*-TeO$_2$ lead to direction-dependent device performances. In the *y* direction, both the *n*-type and *p*-type BL *β*-TeO$_2$ MOSFETs can satisfy the ITRS high-performance (HP) targets with $L_g$ down to 2 nm, superior to the reported oxide semiconductor devices so far (only *n*-type satisfies, as shown in Figure 1(d)). Moreover, the BL *β*-TeO$_2$ NMOS and PMOS exhibit excellent symmetry due to the combined effect of effective mass and UL structure, which can be further optimized by applying tensile strain. As to the *x* direction, we find that the BL *β*-TeO$_2$ MOSFET is applicable to both HP and LP electronics, while the comment 2D TMD (ML MoS$_2$ and MoTe$_2$) and 1D Si devices can be used for only one of them. Hence, our findings shed light on the substantial promise of BL *β*-TeO$_2$ for future CMOS-integrated technology.

## 2. Method

Based on the density functional theory (DFT) and non-equilibrium Green's function (NEGF), we simulate the device characteristics of BL *β*-TeO$_2$ MOSFETs in QuantumATK 2022.[30, 31] An FET is usually comprised of three parts: the source, drain, and channel region. The coupling between the source/drain electrodes and the channel is depicted by the self-energy $\sum_{k_{//}}^{l/r}$, in which *l/r* is the left (source)/right (drain) electrodes, and $k_{//}$ denotes the reciprocal lattice vector parallel to the surface. The broadening matrix $\Gamma_{k_{//}}^{l/r}(E) = i[\sum_{k_{//}}^{l/r} - (\sum_{k_{//}}^{l/r})^\dagger]$ is the imaginary component of $\sum_{k_{//}}^{l/r}$. Thus, the transmission coefficient $T_{k_{//}}(E)$ is calculated by $T_{k_{//}}(E) = Tr[\Gamma_{k_{//}}^{l}(E) G_{k_{//}}(E) \Gamma_{k_{//}}^{r}(E) G_{k_{//}}^\dagger(E)]$, where $G_{k_{//}}(E)$ [$G_{k_{//}}^\dagger(E)$] represents the retarded [advanced] channel Green's function. Averaging $T_{k_{//}}(E)$ over $k_{//}$ in the irreducible Brillouin zone yields the transmission function $T(E)$. Therefore, we can acquire the drain current ($I_{ds}$) according to the Landauer–Büttiker formula:

$$I_{ds} = \frac{2e}{h} \int_{-\infty}^{+\infty} [f_D(E - \mu_D) - f_S(E - \mu_S)] T(E) \, dE \qquad (1)$$

Here, $\mu_D/\mu_S$ and $f_D/f_S$ correspond to the electrochemical potential of drain/source and the Fermi-Dirac distribution functions of drain/source, respectively. The *k*-point meshes are sampled based on the Monkhorst-Pack method by 5×1×164 for both electrodes and channel. The temperature is 300 K, and the PseudoDojo pseudopotential is applied. Neumann, Periodic, and Dirichlet boundary conditions are used for the vertical, transverse, and transport directions, respectively.

We employ the generalized gradient approximation (GGA) exchange-correlation function in the form of Perdew-Burke-Ernzerhof (PBE) in our device simulation.[32-34] In a transistor, there are two kinds of screening effects, the heavily doped carriers and surrounding dielectric environments, which result in the accurate bandgap estimation by the GGA-PBE method. For the doped carriers screening, one example is the degenerately doped ML $MoSe_2$ system possessing a GAA bandgap (1.52 eV) consistent with the GW/experiment bandgaps (1.59/1.58 eV).[35-37] As to the dielectric shielding, previous studies show that the $HfO_2$ sandwiched ML $MoS_2$ system exhibits bandgap renormalization, and the renormalized GW bandgap (1.9 eV) agrees with the GAA bandgap (1.76 eV).[38,39] Besides, the simulated transport characteristics of 5-nm-$L_g$ carbon nanotube (CNT) MOSFETs is in agreement with the experimental results, indicating the viability of the DFT-NEGF method.[40]

## 3. Results

### 3.1 Anisotropic Electronic Structure and On-state Current

The BL $\beta$-$TeO_2$ exhibits an anisotropic atomic structure in the *x* and *y* directions, as displayed in Figure 1(a). The space group symmetry of BL $\beta$-$TeO_2$ is *Pca21* (No. 29). In each layer of the BL $\beta$-$TeO_2$, one oxygen atom is coordinated with two tellurium atoms, while one tellurium atom is coordinated with four oxygen atoms. The optimized lattice parameters of the relaxed BL $\beta$-$TeO_2$ are $a$ = 5.541 Å and $b$ = 5.67 Å, which agree with the previous theoretical results.[41] Figure 1(b) shows the calculated band structure of BL $\beta$-$TeO_2$ at the GGA-PBE level. Both the conduction band minimum (CBM) and valence band maximum (VBM) are located at the Γ point, indicating a direct band gap ($E_g$) of BL $\beta$-$TeO_2$. This band gap, quantified at 2.42 eV, corresponds to the previous theoretical prediction.[41] Owing to the anisotropic atomic structure, the extracted effective masses (*m*) from the band structure also possess anisotropy. The electron/hole *m* ($m_e/m_h$) along the *x* and *y* directions are 7.25/0.69 and 0.36/0.71 $m_0$, respectively; here, $m_0$ is the electron mass.

We next construct the device configuration of BL $\beta$-$TeO_2$ MOSFETs, as shown in Figure 1(c). In this configuration, the source/drain electrodes and the channel are modeled by the highly doped and pristine BL $\beta$-$TeO_2$, respectively. Notably, the channel is separated into two distinct segments: the region beneath the gate and the symmetric underlap (UL) structure. For a MOSFET, one of the crucial parameters to assess the device performance is the on-state current ($I_{on}$), which can be acquired from the $I$-$V$ transfer characteristics at the on-state point ($V_{g,\,on}$, $I_{on}$). $V_{g,\,on}$ is the on-state voltage, and it is defined as $V_{g,\,on} = V_{g,\,off} + V_{dd}$, where $V_{g,\,off}$ is the off-state voltage at the off-state point ($V_{g,\,off}$, $I_{off}$) and $V_{dd}$ is the supply voltage. $I_{off}$ and $V_{dd}$ can be determined by the ITRS criteria. Since the ITRS 2013 and latest International Roadmap for Devices and Systems (IRDS) 2023 versions are only suitable for devices with $L_g$ exceeding 5 and 12 nm,[42] respectively, we extrapolate the ITRS 2013 standard into the sub-5-nm-$L_g$ region based on the existing data above 5 nm, as summarized in Tables S1 and S2.

Owing to the anisotropic structure of BL $\beta$-TeO$_2$, the transport characteristics of corresponding devices also manifest anisotropy. According to previous studies, $I_{on}$ reduces with increasing $m$ when $m < 0.45\ m_0$ because of the reduced electron velocity (inversely proportional to $m$) and increases with enhancing $m$ when $m > 0.45\ m_0$ due to the improved density of states (proportional to $m$).[43-45] For the $n$-type devices, since $m_e$ of BL $\beta$-TeO$_2$ along the $y$ (0.36 $m_0$) and $x$ (7.25 $m_0$) directions are located on different ranges, both these two $m_e$ values possibly lead to a large $I_{on}$. Figure 2(a) shows $I_{on}$ of the BL $\beta$-TeO$_2$ MOSFETs along different transport directions as a function of electron doping concentrations ($1\times10^{13}$, $3\times10^{13}$, and $6\times10^{13}$ cm$^{-2}$) at $L_g = 5$ nm. In fact, $I_{on}$ along the $y$ direction ($y$-$n$) markedly surpass those along the $x$ direction ($x$-$n$) for all the doping concentrations.

To further illustrate the differences between $x$-$n$ and $y$-$n$ BL $\beta$-TeO$_2$ MOSFETs, we plot the conduction band around Γ point and electron transmission coefficients for different transport directions, as presented in Figures 2(b)-(d). The shapes of the conduction band along X-Γ-X' (orange line) and Y-Γ-Y' (blue line) are very similar to each other, which exhibit only one valley. As a result, the electron transmission coefficients along both the $x$ and $y$ directions show one adjacent peak. However, since the conduction band of the X-Γ-X' direction spans a broader $k$ vector window than that of the Y-Γ-Y' direction, the $y$ direction covers wider-range electron transmission coefficients than the $x$-direction counterparts. Hence, $I_{on}$ of the $n$-type BL $\beta$-TeO$_2$ MOSFETs along the $y$ direction exceeds that along the $x$ direction. Through a thorough comparison, we choose $1\times10^{13}$ cm$^{-2}$ as the electron doping concentration for both these two transport directions because it possesses the highest $I_{on}$.

As to the $p$-type devices, the $y$ direction ($y$-$p$) exhibits about two orders of magnitude higher $I_{on}$ than that of the $x$ direction ($x$-$p$) at all the doping concentrations (Figure S1(a)). The impact of effective mass on $I_{on}$ value is relatively minor because of the nearly equivalent $m_h$ for the $y$-$p$ (0.71 $m_0$) and $x$-$p$ (0.69 $m_0$) BL $\beta$-TeO$_2$. Similar to electron transport, we also depict the valence band around the Γ point and hole transmission coefficients (Figures S1(b)-(d)). Around the valence band maximum (VBM), one peak is observed for different *k*-vector directions, resulting in a single peak in the hole transmission coefficients along different transport directions. The transmission coefficient peak in the *y* direction is approximately two orders of magnitude larger than that in the *x* direction. In addition, the *y*-direction transmission coefficient also spans a wider *k*-vector range. These results, which lead to larger $I_{on}$ of the *y-p* BL *β*-TeO$_2$ MOSFETs, are attributed to the more band numbers and broader span of bands around VBM along the X-Γ-X' direction. The doping concentration of $3\times10^{13}$ cm$^{-2}$ is employed for the *y* direction because it exhibits the highest $I_{on}$ compared to the other doping concentrations. Notably, we abandon the *p*-type devices along the *x* direction in subsequent simulations due to their ultra-small $I_{on}$ (< 10 μA/μm).

### 3.2 Mechanism and Effect of Underlap Structure

Based on the doping concentrations obtained above, we simulate the transfer characteristics of the *x-n*, *y-n*, and *y-p* BL *β*-TeO$_2$ MOSFETs, as shown in Figures S2-S4. The extracted $I_{on}$ shows significant improvement by employing the symmetrical UL structure. As an example, we plot $I_{on}$ of the *n*-type BL *β*-TeO$_2$ MOSFETs with/without UL along the *x* and *y* directions in Figures 3(a) and 3(b), respectively. Without UL, both the *x*- and *y*-direction devices fail to meet the ITRS HP goals. By contrast, the UL structure improves $I_{on}$ several times, and thus the $L_g$ scaling limit in terms of ITRS requirements reaches 3 and 2 nm for the devices along the *x* and *y* directions, respectively. To clarify the underlying mechanism of UL in stimulating the device performance, the local density of states (LDOS) and spectrum current of the BL *β*-TeO$_2$ MOSFETs at $L_g$ = 2 nm (UL length of 0 and 2 nm) are depicted in Figure 4. LDOS describes the DOS distribution in real space. The electron barrier *Φ* is deducible from the LDOS by the difference between the conduction band minimum (CBM) of the central channel and the Fermi level of the drain electrode ($\mu_d$). On the other hand, the spectrum current comprises the thermionic current ($I_{therm}$) and tunneling current ($I_{tunnel}$), which is separated by the CBM of the central channel.[46, 47]

At the off-state, the spectrum current is primarily dominated by $I_{tunnel}$ for both UL lengths due to the relatively high *Φ*. Since $I_{tunnel}$ is inversely proportional to *Φ* and barrier width *w* ($I_{tunnel} \propto e^{-w\sqrt{m\Phi}}$),[34, 48] a longer UL length will lead to a larger *w* and hence a smaller $I_{tunnel}$. To realize the same off-state current $I_{off}$ for different UL lengths, *Φ* should be smaller for longer UL (2 nm). Therefore, 0-nm UL exhibits a *Φ* of 0.62 eV, surpassing the 2-nm UL counterparts (0.33 eV). After applying $V_{dd}$, the BL *β*-TeO$_2$ MOSFETs are switched from off-state to on-state. A considerable *Φ* of 0.28 eV is observed for UL = 0 nm because of the large *Φ* at the off-state, which suppresses the presence of $I_{therm}$. In contrast, *Φ* is reduced to -0.05 eV for UL = 2 nm, leading to the domination of $I_{therm}$ over $I_{\text{tunnel}}$. Consequently, the peak spectrum current of 2-nm UL is approximately two orders of magnitude higher than that of 0-nm UL, which in turn results in a larger $I_{\text{on}}$. In a word, a longer UL length leads to a smaller $\Phi$ for both the off-state and on-state, and thus an overall larger on-state current.

Considering the positive effect of UL structure, we derive the UL-optimized $I_{\text{on}}$ of the BL $\beta$-TeO$_2$ MOSFETs along the $x$ and $y$ directions in Figures 3(c) and 3(d), respectively. Due to the anisotropic structure of BL $\beta$-TeO$_2$, the scaling of $I_{\text{on}}$ with $L_{\text{g}}$ is distinct for different transport directions. In the $x$ direction, $I_{\text{on}}$ of the $n$-type BL $\beta$-TeO$_2$ MOSFETs fulfills both the ITRS HP and LP demands at $L_{\text{g}}$ of 3 nm. Moving to the $y$ direction, the $n$-type HP devices are demonstrated to reach the ITRS targets at $L_{\text{g}}$ of 2 nm although the LP transistors cannot (Figure S5). Similar to BL $\beta$-TeO$_2$, previous studies have also revealed the satisfaction of ITRS requirements for various $n$-type oxide semiconductors (ultrathin In$_2$O$_3$, ML TeO$_2$, ML Ga$_2$O$_3$, and BL Ga$_2$O$_3$) transistors at sub-5 nm $L_{\text{g}}$ (Figure 1(d)).[24-27] However, the absence of ultra-short gate length $p$-type devices has limited the realization of CMOS. Fortunately, our discovery reveals that the $p$-type BL $\beta$-TeO$_2$ MOSFETs along the $y$ direction can meet the ITRS standards with $L_{\text{g}}$ down to 2 nm (Figure 3(d)).

Except for the enhancement of $I_{\text{on}}$, the UL structure is also conducive to promoting gate controllability by reducing the impact of the electrodes on channel. As displayed in Figure 4, the variation of $\Phi$ at UL = 2 nm (0.38 eV) surpasses that at UL = 0 nm (0.34 eV), revealing better gate control. The gate controllability is generally described by subthreshold swing $SS = \frac{\partial V_{\text{g}}}{\partial \lg I}$ in the subthreshold region. A smaller $SS$ corresponds to superior gate control. The $SS$ values are significantly reduced with the assistance of UL structure, as displayed in Tables S1 and S2. We thus plot the UL-optimized $SS$ for both HP and LP devices in Figure S6. $SS$ of the $n$-type BL $\beta$-TeO$_2$ MOSFETs along the $y$ direction in all situations exceeds that along the $x$ direction, which is attributed to the relatively smaller $m_{\text{e}}$ of the $y$ direction and hence larger $I_{\text{tunnel}}$ ($I_{\text{tunnel}} \propto e^{-w\sqrt{m\Phi}}$). Notably, the $n$-type and $p$-type devices along the $y$-direction exhibit symmetrical scaling behavior.

## 3.3 Delay Time, Power Dissipation, and Energy-Delay Product

The delay time ($\tau$) and power dissipation (PDP) are two key figures of merits for assessing the FET performance. $\tau$ and PDP reflect the switching speed and energy cost during the off-state and on-state switching processes, respectively. They can be computed by $\tau = C_{\text{t}}V_{\text{dd}}/I_{\text{on}}$ and PDP $= V_{\text{dd}}I_{\text{on}}\tau = C_{\text{t}}V_{\text{dd}}^2$, where $C_{\text{t}}$ is the total capacitance. According to the ITRS criteria, $C_{\text{t}}$ is three times the gate capacitance $C_{\text{g}} = \partial Q_{\text{ch}}/\partial V_{\text{g}}$, in which $Q_{\text{ch}}$ indicates the total charge in the channel. The values of $C_{\text{t}}$ are summarized in Tables S1 and S2. Evidently, both the HP and LP goals outlined by ITRS can be fulfilled by all the $C_t$ of BL $\beta$-TeO$_2$ MOSFETs along different transport directions.

In Figure 5, we depict the UL-optimized $\tau$ and PDP of the BL $\beta$-TeO$_2$ MOSFETs along the $x$ and $y$ directions. Similar to $I_{on}$ and SS, the anisotropic scaling behavior is also discovered. For the $x$ direction, both the HP and LP devices satisfy the ITRS demands at $L_g$ of 1-3 nm for $\tau$ and PDP (Figures 5(a) and 5(c)). As to the $y$ direction, the $L_g$ scaling limit of the $n$-type and $p$-type BL $\beta$-TeO$_2$ MOSFETs in terms of ITRS HP requirements reaches 1 nm (Figures 5(b) and 5(d)). Moreover, $\tau$ and PDP values of the $n$-type devices at a given $L_g$ are comparable with those of the $p$-type devices, revealing good NMOS and PMOS symmetry. Finally, we also found that the BL $\beta$-TeO$_2$ MOSFETs along the $y$ direction are applicable for the LP electronics with $L_g$ decreased to 2 nm, as shown in Figure S7.

Searching for a balance between $\tau$ and PDP is very crucial for the operation of a transistor. The energy-delay product (EDP) is usually employed to describe such balance. EDP can be acquired by the formula EDP = $\tau \times$ PDP. Figure 6 shows the EDP values of various BL $\beta$-TeO$_2$ MOSFETs at $L_g$ = 1-3 nm. The comment 2D material (ML MoS$_2$ and MoTe$_2$) MOSFETs are also shown for comparison.[49, 50] BL $\beta$-TeO$_2$ is superior to the ML MoS$_2$ and MoTe$_2$ counterparts for the $n$-type HP, $n$-type LP, and $p$-type HP devices in terms of EDP. As to the $p$-type LP devices, EDPs of BL $\beta$-TeO$_2$ MOSFETs are better than those of the ML MoTe$_2$ counterparts, while they are comparable with the ML MoS$_2$ counterparts. Overall speaking, the BL $\beta$-TeO$_2$ MOSFETs exhibit better EDP performance than the comment ML MoS$_2$ and MoTe$_2$ devices.

### 3.4 Symmetric NMOS and PMOS performance

To achieve better logic operation of CMOS, the symmetry of $n$-type and $p$-type transistors should preferably use the same channel material. The single-channel-material system is beneficial for simplifying the manufacturing process of circuits. For the oxide-semiconductor-based CMOS, many efforts have been put into realizing symmetric NMOS and PMOS, such as $n$-type SnO/In$_2$O$_3$/IGZO and $p$-type SnO.[51-53] However, all these studies only focus on devices with long channel lengths (usually on the micron scale). The lack of $p$-type oxide semiconductor transistors at the short-channel region (< 10 nm) renders the implementation of ultra-short CMOS a challenging problem. The above results have revealed the potential of BL $\beta$-TeO$_2$ along $y$ direction in ultra-short CMOS design. It is worth further exploring whether the single-channel-material NMOS and PMOS symmetry can be achieved in the BL $\beta$-TeO$_2$ system.

There are two factors, namely the effective mass $m$ and UL structure, that will affect the symmetry of BL $\beta$-TeO$_2$ MOSFETs. On the one hand, $m_e$ (0.36 $m_0$) and $m_h$ (0.71 $m_0$) of the BL $\beta$-TeO$_2$ are distributed on both sides of 0.45 $m_0$, which will probably lead to a relatively symmetric performance. On the other hand, the UL structure is conducive to optimizing the device performance and thus will also influence the symmetry. To elaborate on this, we plot the NMOS and PMOS ratios at the same $L_g$ and UL lengths in Figures 7(a) and 7(b). For $I_{on}$ and $SS$ (Figure 7(a)), the *p*-type BL *β*-TeO$_2$ MOSFETs are hard to switch off at $L_g$ of 1-3 nm without UL, which results in the absence of both ratios. When adding UL (1 and 2 nm), the ratio appears, and most of them are located in the 0-2 range (except for the $I_{on}$ ratio at $L_g/L_{UL}$ = 3/2 nm). Especially, the ratios of $SS$ fall between 1.16 and 1.26 for all the UL lengths. As to $τ$ and PDP (Figure 7(b)), the ratios at 1- and 2-nm-UL are ranged in 0-2. Those results reveal good NMOS and PMOS symmetry of BL *β*-TeO$_2$ along the *y* direction.

Strain engineering has been demonstrated to be an effective method for boosting the device performance.[34, 54, 55] To further optimize the NMOS and PMOS symmetry of the BL *β*-TeO$_2$ MOSFETs, we modulate *m* of BL *β*-TeO$_2$ by applying the tensile strain. As an example, only the 2% tensile strain is tested. The variation of $m_e/m_h$ ratio under tensile strain is displayed in Figure 7(c). We can find the tensile strain enhances the $m_e/m_h$ ratios from 0.50 to 0.58, which is closer to 1. To examine the variation in corresponding device performance, we simulated the transfer curves of the strained BL *β*-TeO$_2$ MOSFETs at $L_g$ = 3 nm and $L_{UL}$ = 2 nm. The extracted $I_{on}$ ratio of NMOS and PMOS is plotted in Figure 7(d). Owing to the increase in $m_e/m_h$ ratio, the $I_{on}$ ratio is decreased from 3.3 (without strain) to 1.5 (tensile strain). Therefore, the NMOS and PMOS symmetry can be further improved by applying the tensile strain.

## 4. Discussion

We now discuss the potential experimental realization of BL *β*-TeO$_2$ MOSFETs in terms of the fabrication process. The key fabrication steps are illustrated in Figure 8(a). First, gate metal electrodes can be deposited through standard lithography and liftoff techniques followed by gate dielectric growth with atomic layer deposition or sputtering (Steps 1 and 2). This will form the bottom gate layer. Next, the active BL TeO$_2$ channel is deposited (Step 3). Currently, liquid metal-based synthesis has emerged as a promising technique to obtain ultrathin TeO$_2$. This synthesis exploits the self-limiting Cabrera-Mott surface oxidation in a roll-transfer process.[56] By rolling a molten eutectic droplet of ~95% Se and 5% Te across target substrates, ultrathin TeO$_2$ films can be delaminated from the liquid metal droplet surface to form the active channel. Next, metals with appropriate work function alignment are deposited as source-drain contacts with standard lithography and liftoff (Step 4). Finally, the top dielectric layer can be grown via atomic layer deposition,[57] before deposition of the top metal gate electrode to form the top gate layer (Step 5).

A crucial consideration for the BL TeO$_2$ synthesis is substrate surface uniformity. As liquid metal oxide printing is a conformal process, the target substrate surface should ideally be smooth to avoid strain or, worse, cracks and wrinkles.[58] However, this may introduce additional fabrication complexities and limitations as the bottom gate electrode presents a finite thickness. Here, Pt and Pd may be useful to realize sub-5 nm thick continuous electrodes from evaporation, but even 5 nm thick electrodes can introduce unacceptable roughness when printing ~1.5 nm thick TeO$_2$. Few-layer graphene can also be used as electrodes but its integration and uniformity control may be challenging to scale.[59] Alternatively, mesa structures shown in Figure 8(b) can be created through chemical-mechanical polishing and metal etching rather than liftoff processes.

Currently, the experimental realization of ultrathin TeO$_2$ is through manual roll transfer of liquid metal droplets using a glass rod where film sizes are limited to mm scale.[60] For scalability, roll-to-roll or squeeze printing can potentially be used, but the melting point of Te/Se eutectic mixture poses a challenge (synthesis performed at ~270 °C). Alternative eutectic alloys with lower melting points may be useful to address this. After TeO$_2$ synthesis, appropriate selective etching strategies will need to be developed to pattern desired channel geometries. To realize efficient *n*-type or *p*-type FETs, appropriate alignment of metallic contact work functions with TeO$_2$ band edges is necessary to realize efficient carrier injection.[16] However, more work is required for contact engineering of ultrathin TeO$_2$ where only *p*-type FETs have been demonstrated.[29, 61] An interesting approach is to pre-fabricate contact electrodes on the substrate before printing the active TeO$_2$ channel, which can be beneficial for scalability and process integration [Figure 8(c)].

## 5. Conclusion

In conclusion, we investigate the transport characteristics of sub-5-nm $L_g$ BL *β*-TeO$_2$ MOSFETs by combining DFT and NEGF methods. Owing to the anisotropic electronic structure of BL *β*-TeO$_2$, the corresponding device performance also manifests anisotropy. In the *y* direction, the BL *β*-TeO$_2$ is superior to the reported oxide semiconductors because not only the *n*-type device but also the *p*-type device could fulfill the ITRS HP criteria at sub-5 nm $L_g$. Besides, we find that the combined effect of effective mass and UL structure results in the symmetric scaling behavior of BL *β*-TeO$_2$ NMOS and PMOS, which is presented for the first time in the ultra-short oxide semiconductor FETs. For the *x* direction, the *n*-type 3-nm-$L_g$ BL *β*-TeO$_2$ MOSFETs are shown to be suitable for both the HP and LP electronics in terms of ITRS standards. Hence, BL *β*-TeO$_2$ is a very promising channel candidate for next-generation CMOS applications.

## Data Availability Statement

The data that support the findings of this study are available from the corresponding authors upon reasonable request.

## Conflict of Interest

The authors declare no conflict of interest.

## Acknowledgment

This work is supported by the Ministry of Science and Technology of China (No.2022YFA1203904 and No. 2022YFA1200072), the National Natural Science Foundation of China (No. 91964101, No. 12274002, and No. 12164036, No. 62174074), the China Scholarship Council, the Foundation of He'nan Educational Committee (Grant No. 23A430015), the Fundamental Research Funds for the Central Universities, the Natural Science Foundation of Ningxia of China (No. 2020AAC03271), the youth talent training project of Ningxia of China (2016), and the High-performance Computing Platform of Peking University. Y. S. A. and Linqiang Xu acknowledge the support of Singapore University of Technology and Design Kickstarter Initiatives (SKI) under the Award No. SKI 2021_01_12. Y.S.A. is also supported by SMU-SUTD Joint Grant (Award No. 22-SIS-SMU-054) and SUTD-ZJU Thematic Research Grant (Award No. SUTD-ZJU (TR) 202203). The computational work for this article was partially performed on resources of the National Supercomputing Centre, Singapore (https://www.nscc.sg). C.S. Lau acknowledges support by the Agency for Science, Technology, and Research (A*STAR) under its MTC YIRG grant No. M21K3c0124 and MTC IRG grant No. M23M6c0103.

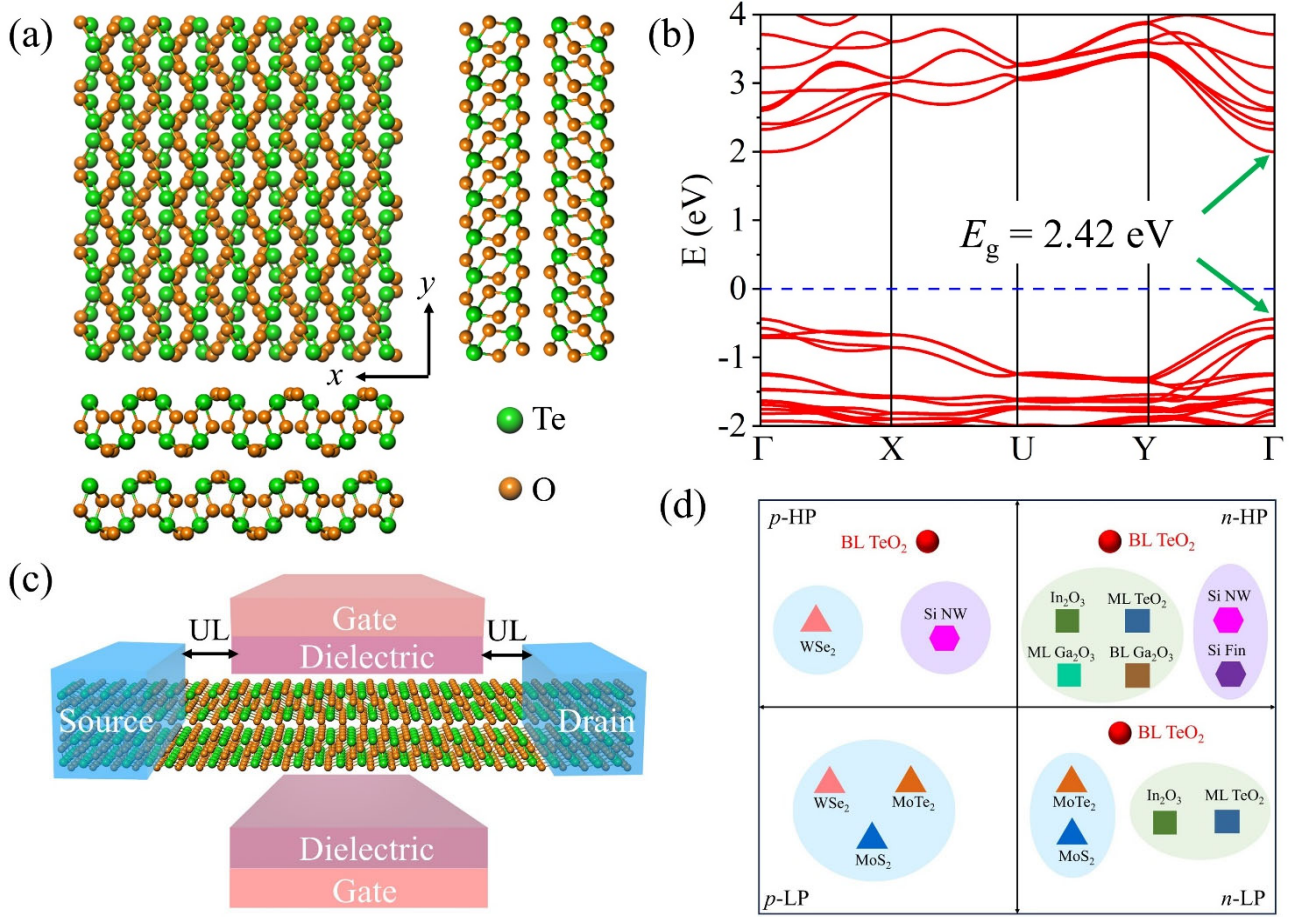

**Figure 1.** (a) Top and side views of the BL $\beta$-TeO$_2$ structure. (b) Calculated band structure of the BL $\beta$-TeO$_2$ at the GGA-PBE level. (c) Schematic diagram of the BL $\beta$-TeO$_2$ MOSFET. (d) Comparison between BL $\beta$-TeO$_2$ and other oxide material, 2D TMD material, and 1D Si material MOSFETs that satisfy ITRS criteria at sub-5-nm $L_g$ for different types ($n$ and $p$) and different applications (HP and LP).[24-27, 49, 50, 62-64]

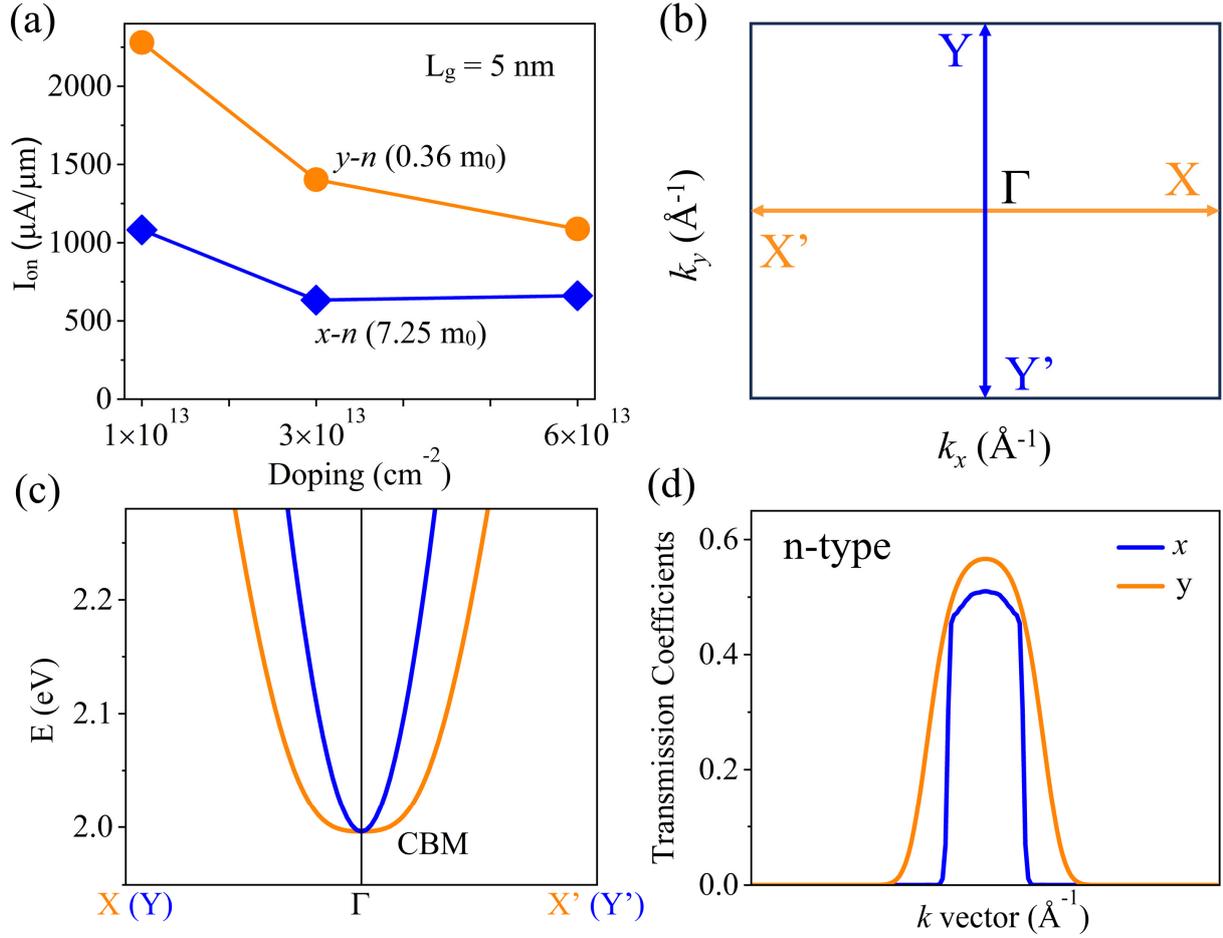

**Figure 2.** (a) $I_{on}$ versus doping concentrations for *n*-type MOSFETs along different transport directions at $L_g$ = 5 nm. The electron effective masses are given in parentheses. (b) Schematic diagram of the Brillouin zone for the BL *β*-TeO$_2$. (c) Conduction band of the BL *β*-TeO$_2$ along the X (0.5, 0, 0)-Γ (0, 0, 0)-X' (-0.5, 0, 0) and Y (0, 0.5, 0)-Γ (0, 0, 0)-Y' (0, -0.5, 0) in the Brillouin zone. (d) Electron transmission coefficients of the *n*-type BL *β*-TeO$_2$ versus *k* vector at the on-state.

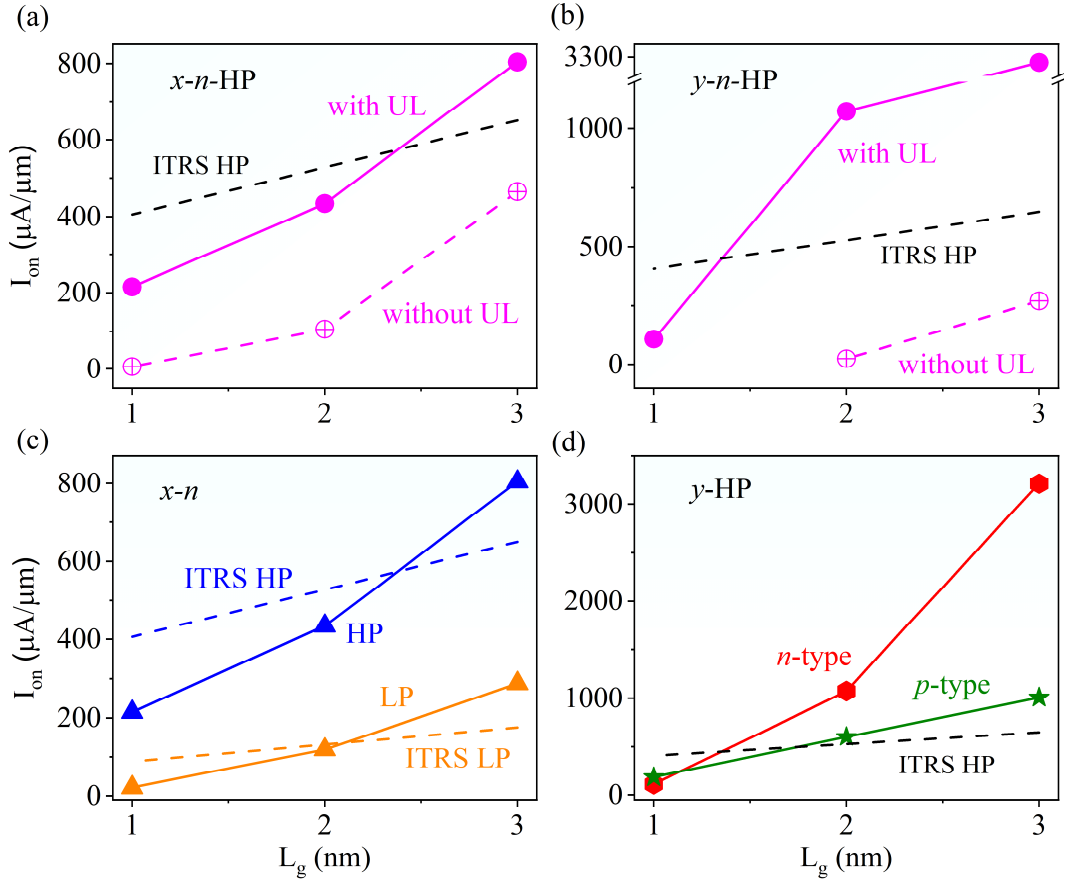

**Figure 3.** $I_{on}$ of the $n$-type BL $\beta$-TeO$_2$ MOSFETs versus $L_g$ along the (a) $x$ direction and (b) $y$ direction for the HP applications. The data with and without UL structure are indicated by different purple circles. (c) UL-optimized $I_{on}$ of $n$-type BL $\beta$-TeO$_2$ MOSFETs along the $x$ direction versus $L_g$ for both the HP and LP applications. (d) UL-optimized $I_{on}$ of $n$-type and $p$-type BL $\beta$-TeO$_2$ MOSFETs along the $y$ direction as a function of $L_g$ for the HP applications. The dashed lines in (c) and (d) represent the ITRS standards.

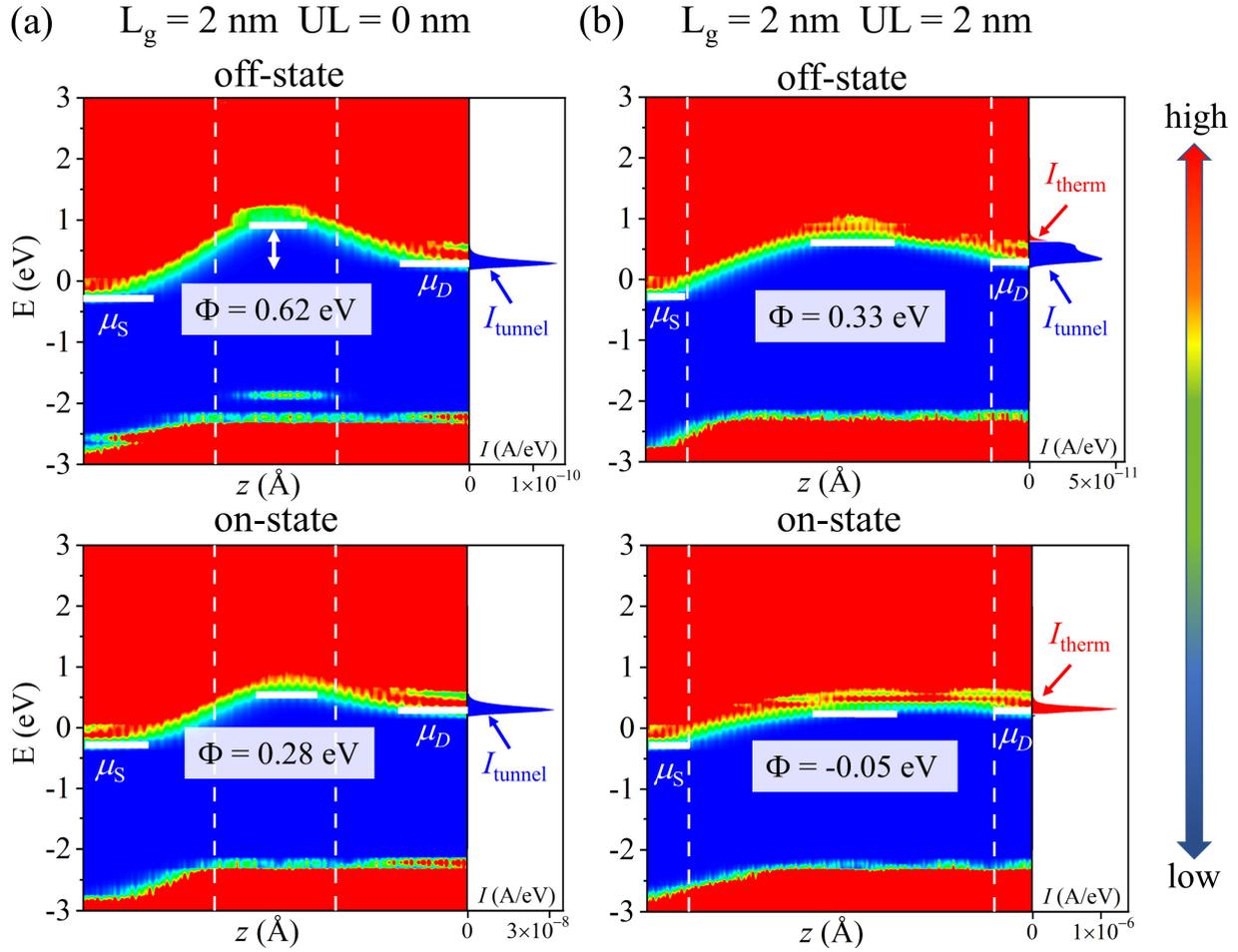

**Figure 4.** LDOS and spectrum current of the *n*-type BL *β*-TeO$_2$ MOSFETs along *y* direction at (a) UL = 0 nm and (b) UL = 2 nm of $L_g$ = 2 nm. *Φ*, *μ*$_s$, and *μ*$_d$ indicate the electron barrier height, the Fermi level of source, and the Fermi level of drain, respectively. The electrode and channel regions are separated by the white dashed line.

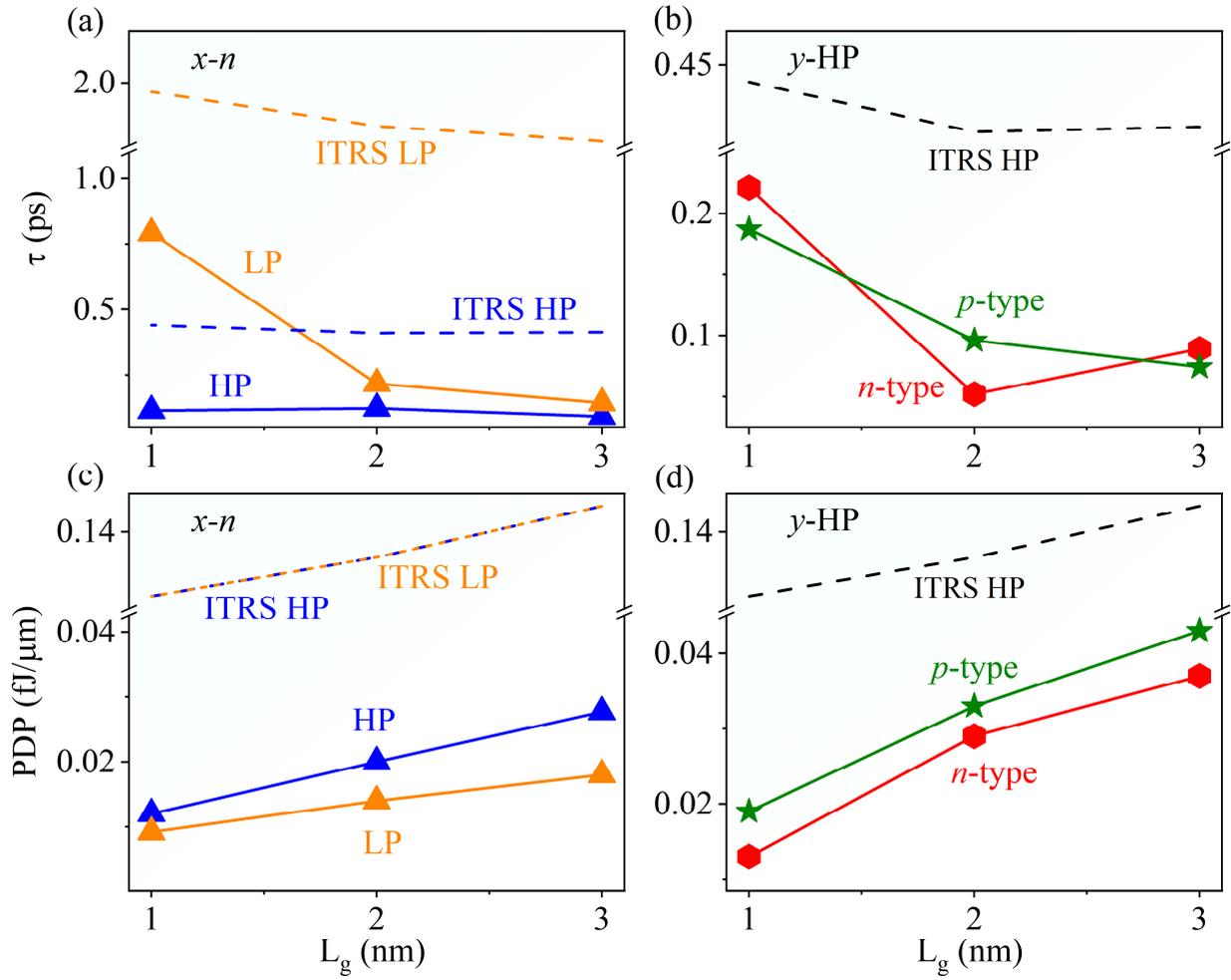

**Figure 5.** UL-optimized $\tau$ versus $L_g$ along (a) $x$ direction and (b) $y$ direction for BL $\beta$-TeO$_2$ MOSFETs. UL-optimized PDP versus $L_g$ along (c) $x$ direction and (d) $y$ direction for BL $\beta$-TeO$_2$ MOSFETs. The dashed lines indicate the ITRS requirements.

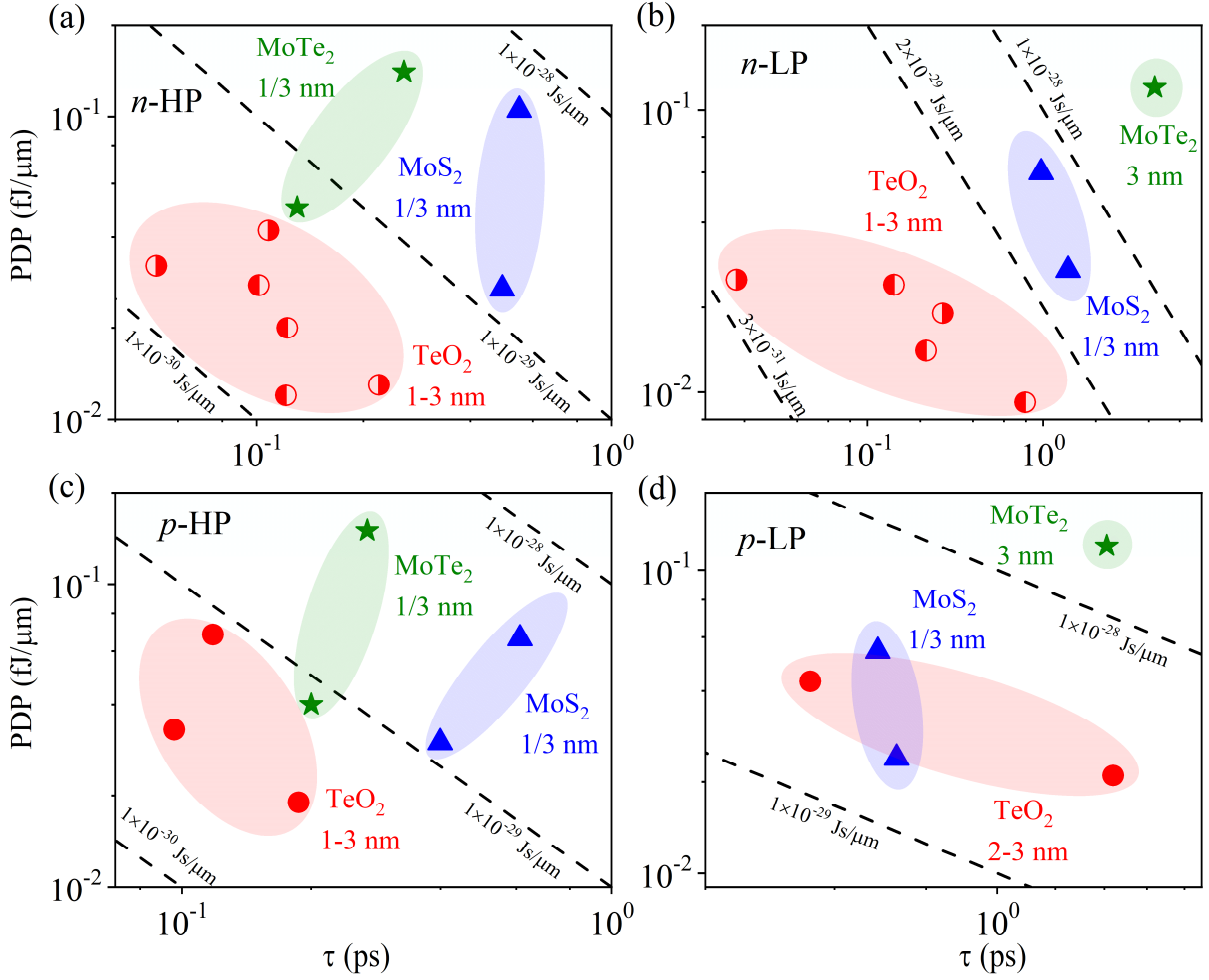

**Figure 6.** Optimal PDP versus $\tau$ for (a) *n*-type HP, (b) *n*-type LP, (c) *p*-type HP, and (d) *p*-type LP MOSFETs with $L_g$ of 1-3 nm. In (a) and (b), the *n*-type BL $\beta$-TeO$_2$ MOSFETs along the *x* and *y* directions are depicted by the right and left self-hollow red circles, respectively. In (c) and (d), the *p*-type BL $\beta$-TeO$_2$ MOSFETs only along the *y* direction. The black dashed lines show the values of energy-delay product EDP = $\tau \times$ PDP. The data of ML MoS$_2$ and MoTe$_2$ are computed by the quantum transport simulations at $V_{dd}$ = 0.64 V.[49, 50]

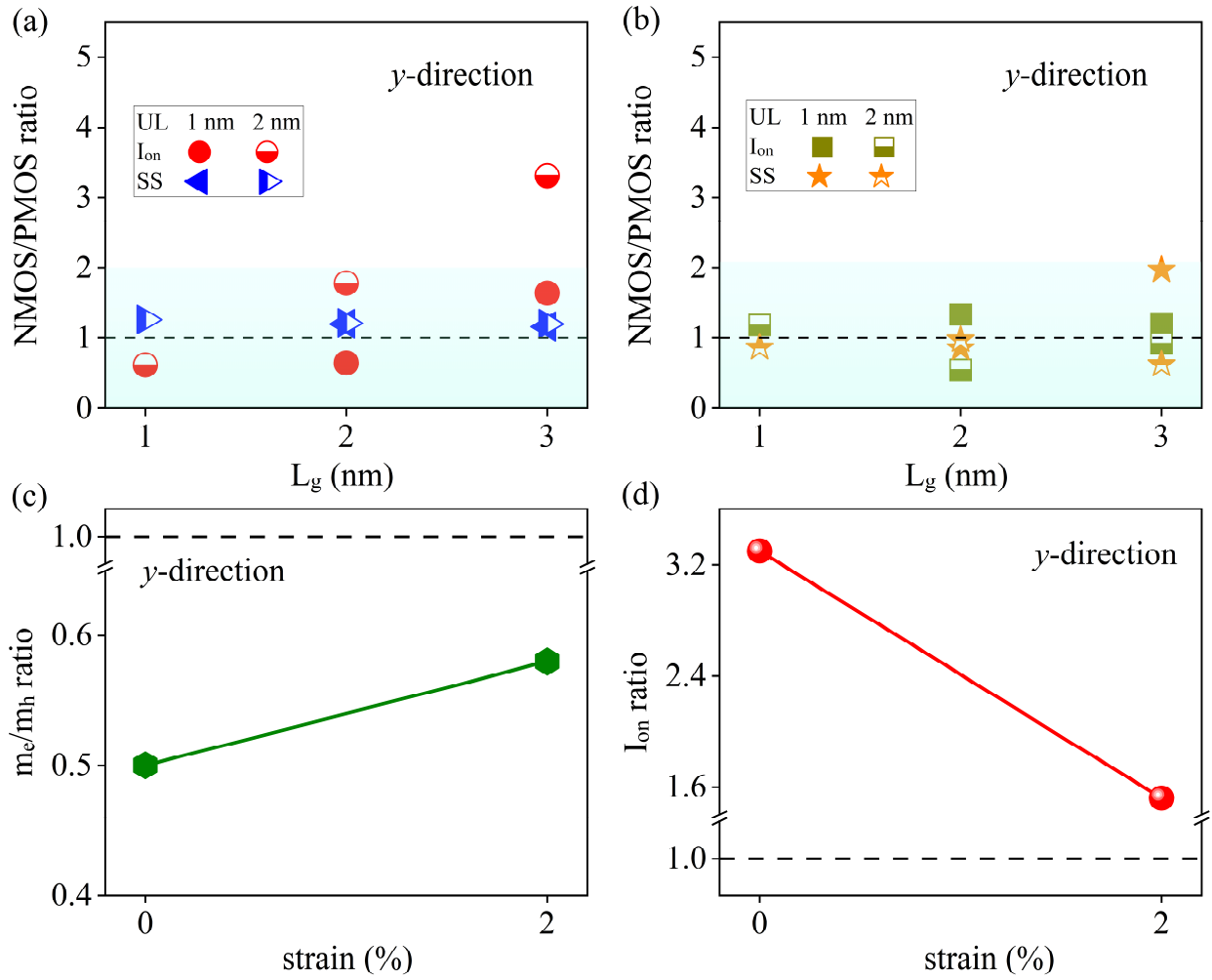

**Figure 7.** NMOS and PMOS ratio of (a) $I_{on}$ and $SS$, (b) $\tau$ and PDP for the BL $\beta$-TeO$_2$ MOSFETs along $y$ direction. The ratio is calculated at the same gate and UL lengths for NMOS and PMOS devices. (c) $m_e/m_h$ ratio as a function of the strain. As an example, we only try 2% tensile strain. (d) $I_{on}$ ratio of BL $\beta$-TeO$_2$ versus strain at $L_g$ = 3 nm and $L_{UL}$ = 2 nm.

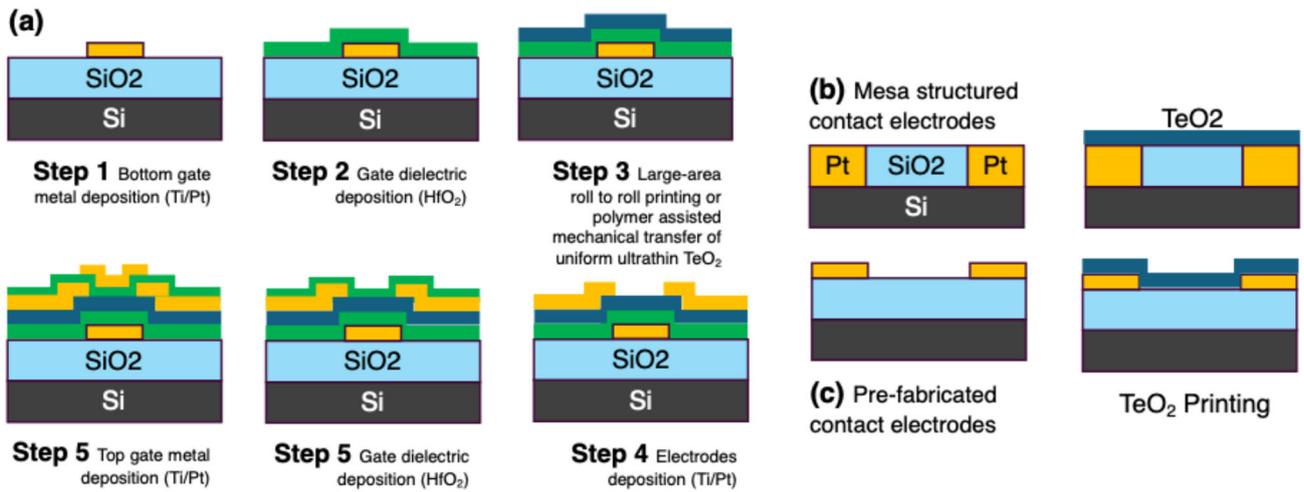

**Figure 8.** Potential experimental fabrication processes of BL *β*-TeO$_2$ MOSFET. (a) BL *β*-TeO$_2$ transistors can be fabricated based on 5 key steps. (b) and (c) show mesa-structured electrodes and pre-fabricated metal contacts which may be further employed to improve the quality of BL *β*-TeO$_2$ MOSFET.

**TOC**

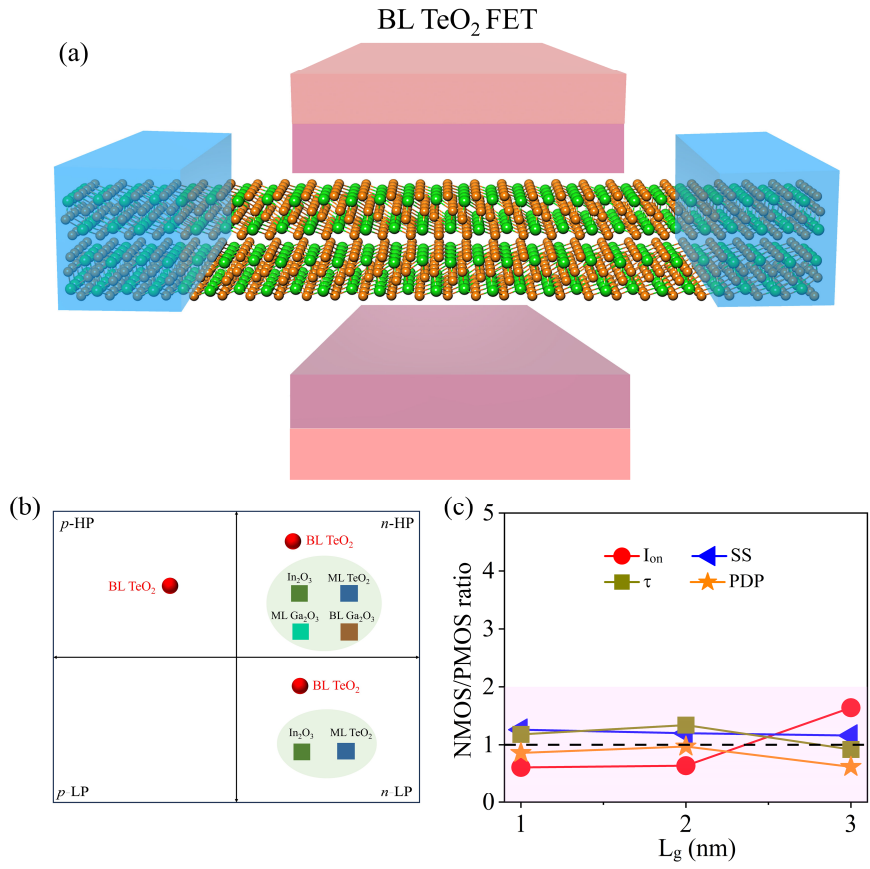